\documentclass{iau}
\usepackage{graphicx}

\title[Tracing GC internal dynamical evolution with BSS] 
{Exotic populations in Globular Clusters: Blue Stragglers as tracers of
the internal dynamical evolution of stellar systems}

\author[F.R. Ferraro]   
{Francesco R. Ferraro$^1$
}

\affiliation{$^1$Department of Physics and Astronomy, University of Bologna, \,Viale Berti Pichat 6/2,
40127- Bologna, Italy \,email: {\tt francesco.ferraro3@unibo.it} \\[\affilskip]
 }

\pubyear{2015}
\volume{312}  
\pagerange{...-....}
\jname{Star Clusters and Black Holes in Galaxies across cosmic time}
\editors{A.C. Editor, B.D. Editor \& C.E. Editor, eds.}
\begin{document}

\maketitle

\begin{abstract}
In this paper I  present an overview of the main observational  properties of a
special class of exotic objects (the so-called  Blue Straggler Stars, BSSs) in Galactic
Globular Clusters (GCs).  The BSS specific frequency and their radial
distribution  are discussed in the framework of using this stellar
population as probe of GC internal dynamics. In particular, the shape
of the BSS radial distribution has been found to be a powerful tracer
of the dynamical evolution of stellar systems, thus allowing the definition
of an empirical ``clock''able to measure the dynamical age 
of stellar aggregates from pure observational properties.

\end{abstract}

\firstsection 
\section{Introduction}

Globular clusters (GCs) are the only cosmic structures 
that within a time-scale shorter than the age of the Universe
undergo nearly all the physical processes known in stellar dynamics
(\cite[Meylan \& Heggie 1997]{mh97}). Gravitational interactions and collisions among single
stars and/or binaries are expected to be quite frequent, especially in high
density environments  (e.g. \cite[Hut et al. 1992]{hut92}) characterizing the inner regions of GCs. 
Such a pronounced dynamical activity can generate
populations of exotic objects, like X-ray binaries, millisecond
pulsars and blue straggler stars (BSSs; see, e.g.,
\cite[Paresce et al. 1992]{paresce92},
\cite[Bailyn 1995]{bailyn95},
\cite[Bellazzini et al. 1995]{bellazzini95},
\cite[Ferraro et al. 2001]{fer01},
\cite[Ransom et al. 2005]{ransom05},
\cite[Pooley \& Hut 2006]{pooleyhut06},
\cite[Ferraro et al. 2009a]{fer09ter},
\cite[Lanzoni et al. 2010]{lan10}).
 
Blue straggler stars (BSSs) are commonly defined as stars brighter and
bluer than the main-sequence (MS) turnoff in the host stellar cluster.
They are thought to be central H-burning stars, more massive
than the MS turnoff stars 
 (\cite[Shara et al. 1997]{shara97},
 \cite[Gilliland et al. 1998]{gilli98},
 \cite[De Marco et al. 2004]{de04},
 \cite[Fiorentino et al. 2014]{fiore14}).  In stellar
  systems with no evidence of recent star formation, their origin
  cannot be explained in the framework of normal single-star evolution.
  Two main formation channels are currently favored: (1) mass
  transfer (MT) in binary systems (\cite[McCrea 1964]{mc64})  possibly up to the complete
  coalescence of the two stars, and (2) stellar collisions (\cite[Hills \& Day 1976]{hills76}).
Both these processes can potentially bring new hydrogen into the core
and therefore ``rejuvenate'' a star to its MS stage (e.g., \cite[Lombardi et al. 1995]{lom95}, \cite[Lombardi et al 2002]{lom02}; 
\cite[Chen \& Han 2009]{chen09}).
 
MT in binaries might be the dominant formation channels in
all environments (e.g., \cite[Knigge et al. 2009]{knigge09}; \cite[Leigh et al. 2013]{leight13}), and
most likely it is so in low-density GCs, open clusters and the
Galactic field (\cite[Ferraro et al. 2006a]{fer06ome}; \cite[Sollima et al. 2008]{soll08}; 
\cite[Mathieu et al. 2009]{mathieu09}; \cite[Gosnell et al. (2014)]{Gos14}; \cite[Preston \& Sneden 2000]{pre00}).  Collisions are believed to be
important especially in dense environments, such as the cores of
globular clusters (GCs; \cite[Bailyn 1992]{bai92}; \cite[Ferraro et al 1993]{fer93},\cite[Ferraro et al 1997]{fer97}
\cite[Ferraro et al 2003a]{fer03six}) and even the center of some open clusters 
(\cite[Leonard \& Linnell 1992]{leli92}; \cite[Glebbeek et al. 2008]{gle08}).

\begin{figure}[!t]
\begin{center} 
\includegraphics[scale=0.35]{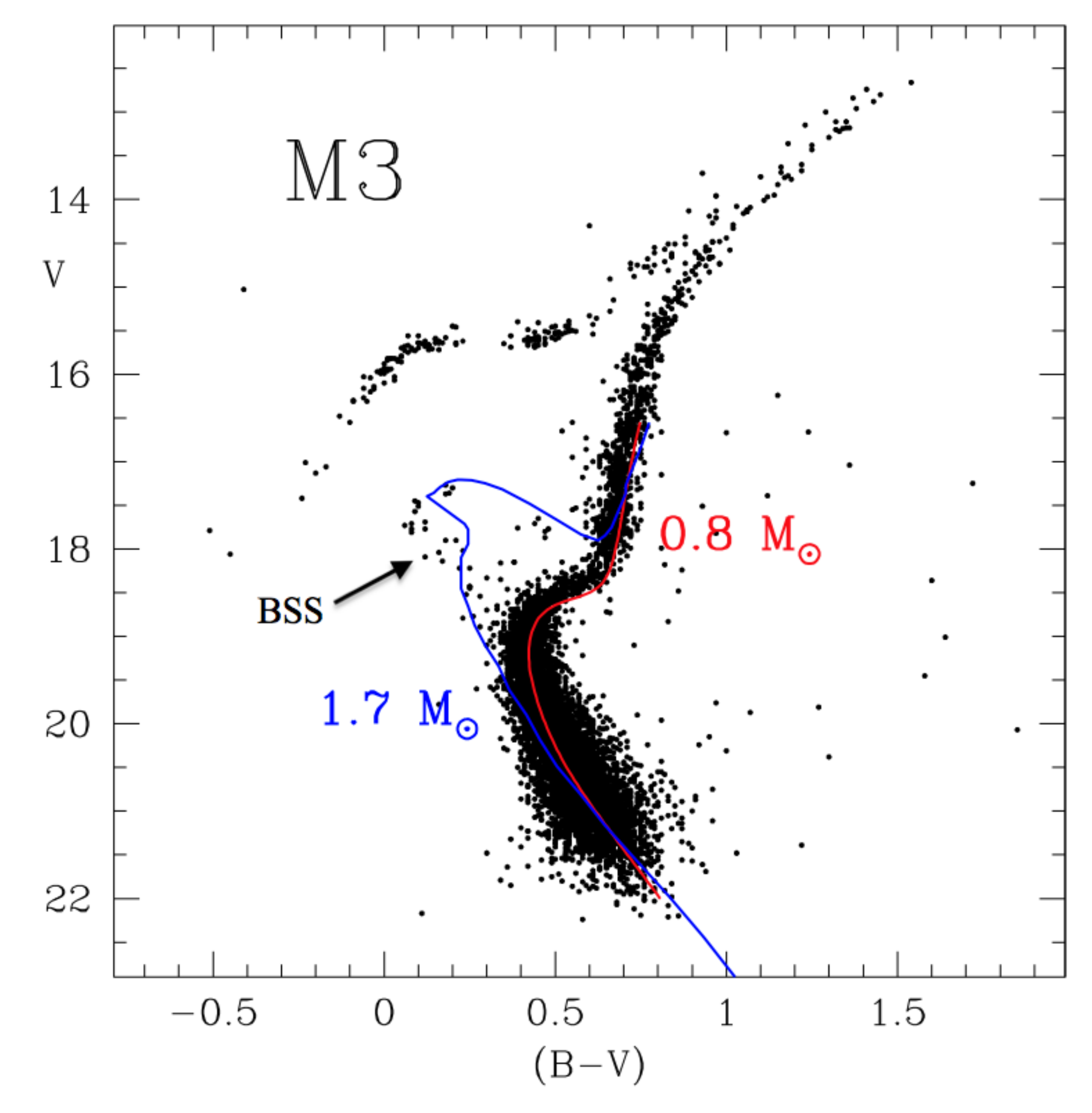}
\caption{\footnotesize Optical CMD of the globular cluster M3, with the location of
  BSSs indicated by the arrow. The theoretical track corresponding to
  $0.8 M_\odot$ well reproduces the main evolutionary sequences of the
  cluster, while BSSs populate a region of the CMD where core
  hydrogen-burning stars of $\sim 1.7 M_\odot$ are expected.
  From \cite[Buonanno et al. (1994)]{buo94}
  }
\label{m3}
\end{center}
\end{figure}
 
Because of their large number of member stars, GCs are the ideal
environment for BSS studies. Nominally all the GCs observed so far
have been found to harbor a significant number of BSSs 
(\cite[Piotto et al. 2004]{pio04}; \cite[Leigh et al. 2007]{leigh07}).  
Moreover, the stellar density in GCs
varies dramatically from the central regions to the outskirts, and
since BSSs in different environments (low- versus high-density) could
have different origins 
(e.g., \cite[Fusi Pecci et al. 1992]{fp92}; \cite[Ferraro et al. 1995]{fer95},\cite[Davies et al. 2004]{davies04}), 
these stellar systems allow to investigate both formation
channels simultaneously. However a clear distinction is hampered by
the internal dynamical evolution of the parent cluster 
(\cite[Ferraro et al. 2012]{fer12}). In fact, having masses larger than normal cluster stars,
BSSs are affected by dynamical friction, a process that drives the
objects more massive than the average toward the cluster center, over
a timescale which primarily depends on the local mass density (e.g.,
\cite[Alessandrini et al. 2014]{ale14}).  Hence, as the time goes on, heavy objects
(like BSSs) orbiting at larger and larger distances from the cluster
center are expected to drift toward the core: as a consequence, the
radial distribution of BSSs develops a central peak and a dip, and the
region devoid of these stars progressively propagates outward.

\cite[Ferraro et al. (2012)]{fer12} used this argument to define the so-called
``dynamical clock'', an empirical tool able to measure the dynamical
age of a stellar system from the shape of its BSS radial
distribution. This appears indeed to provide a coherent interpretation
of the variety of BSS radial distributions observed so far: GCs with a
flat BSS radial distribution 
(\cite[Ferraro et al. 2006b]{fer06ome}; \cite[Dalessandro et
al. 2008a]{dale2419}; \cite[Beccari et al. 2011]{bec11}) 
are dynamically young systems, GCs
with bimodal distributions (e.g., \cite[Ferraro et al. 1993]{fer93}, 
\cite[Ferraro et al. 2004]{fer04}; 
\cite[Lanzoni et al. 2007a]{lan07}; 
\cite[Dalessandro et al. 2008b]{dale6388}; \cite[Beccari et al. 2013]{bec13}, and
references therein) are dynamically intermediate-age systems (their
actual dynamical age being determined by the distance of the dip of the
distribution from the cluster center), and GCs with a single-peaked
BSS distribution (\cite[Ferraro et al. 1999a]{fer99m80}; \cite[Lanzoni et al. 2007b]{lan071904};
\cite[Contreras~Ramos et al. 2012]{contre12}; \cite[Dalessandro et al. 2013]{dale13}) are dynamically
old systems.

 \begin{figure}[!t]
\begin{center} 
\includegraphics[scale=.55]{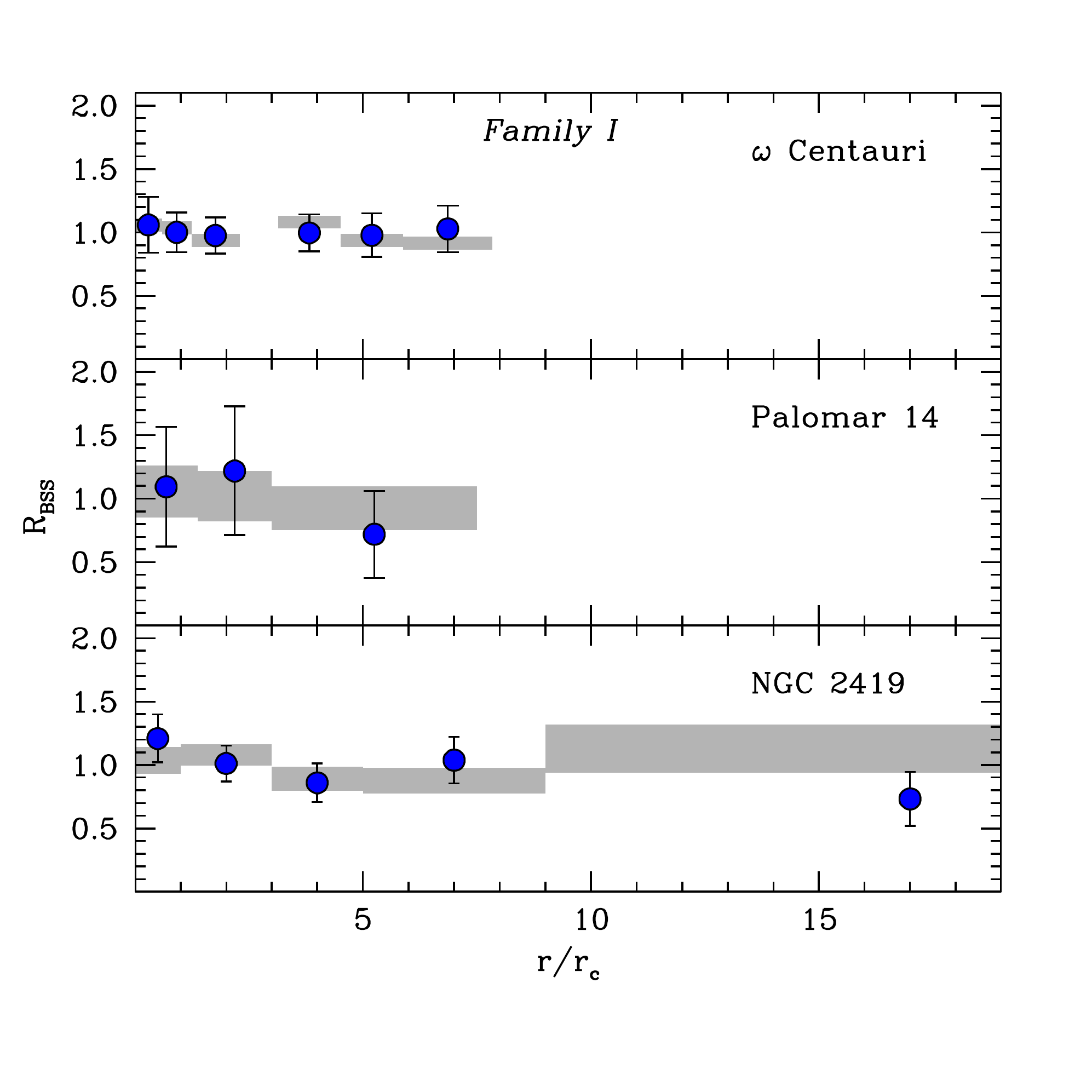}
\caption{\footnotesize BSS radial distribution observed in $\omega$\,Centauri,
  Palomar\,14 and NGC\,2419, with the large filled circles marking the values
  of the double normalized specific frequency $R_{\rm BSS}$. The distribution of the
  double normalized ratio measured for RGB or HB stars is also shown
  for comparison (grey strips). The BSS radial distribution is
  flat and totally consistent with that of the reference population,
  thus indicating a low degree of dynamical evolution for these three
  GCs (\emph{Family I}). From \cite[Ferraro et al. (2012)]{fer12}. }
\label{dyncl1}
\end{center}
\end{figure}

\begin{figure}[!t]
\begin{center} 
\includegraphics[scale=.55]{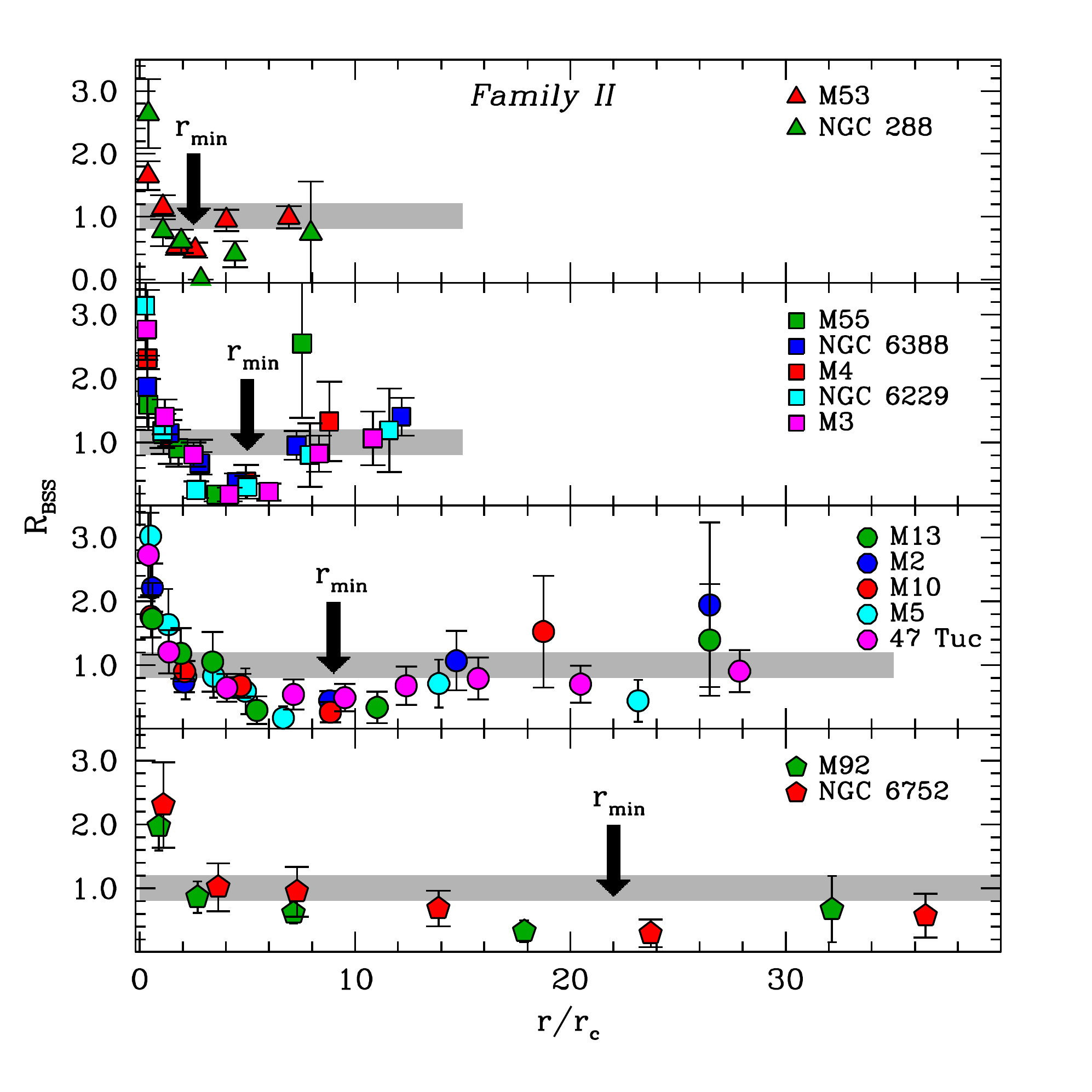}
\caption{\footnotesize BSS radial distribution observed in clusters of intermediate
  dynamical age (\emph{Family II}). The distribution is clearly
  bimodal and the radial position of the minimum (marked with the
  arrow and labelled as $r_{\rm min}$) clearly moves outward from top
  to bottom, suggesting that the bottom clusters are more dynamically
  evolved than the upper ones.  For the sake of clarity, the grey
  bands schematically mark the distribution of the reference
  populations.From \cite[Ferraro et al. (2012)]{fer12}. }
\label{dyncl2}
\end{center}
\end{figure}

\begin{figure}[!t]
\begin{center} 
\includegraphics[scale=.55]{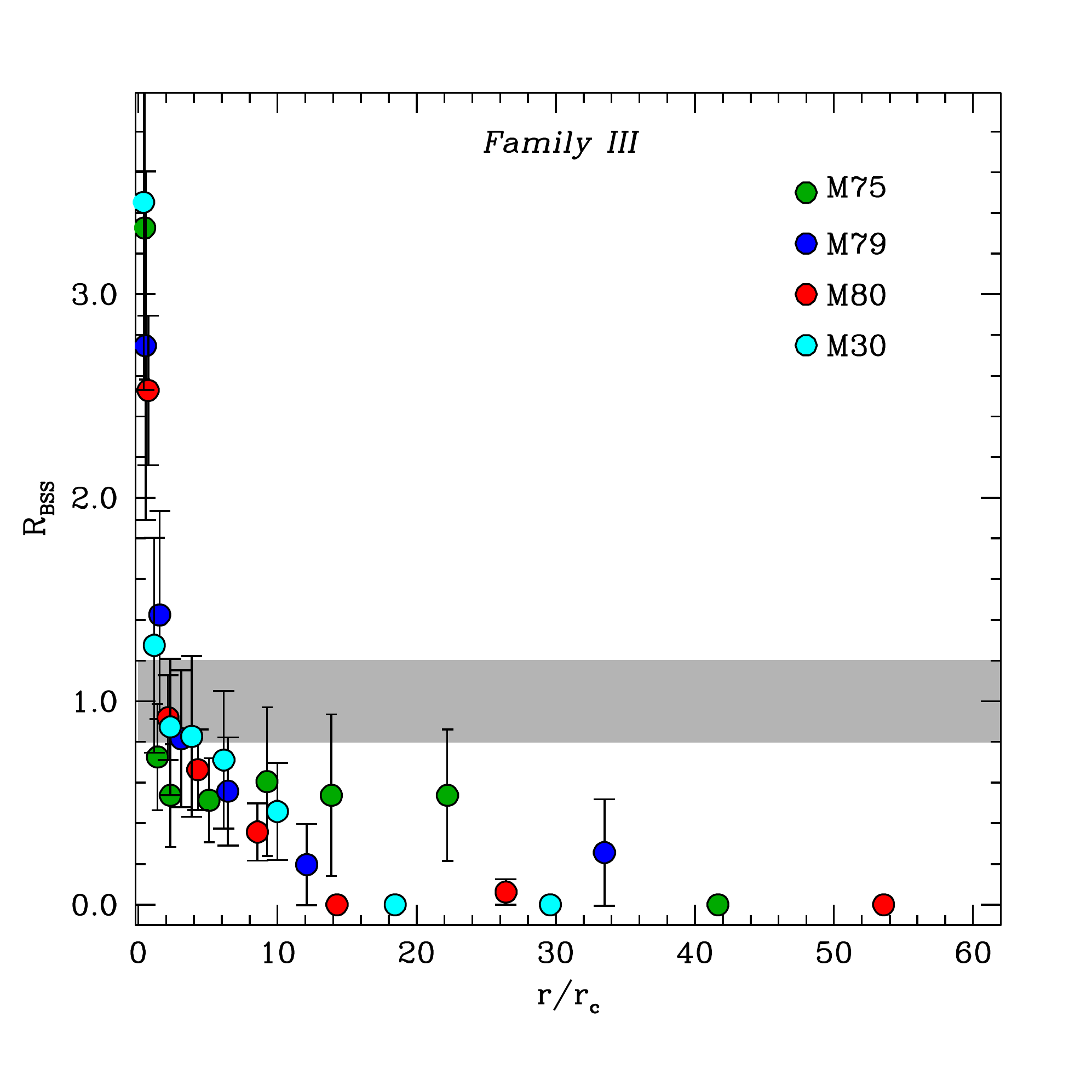}
\caption{\footnotesize BSS radial distribution for dynamically old clusters
  (\emph{Family III}): only a central peak is visible, while the
  external upturn is no more present because of the dynamical friction action out
  to the
  cluster outskirts. From \cite[Ferraro et al. (2012)]{fer12}}
\label{dyncl3}
\end{center}
\end{figure}

\begin{figure}[!t]
\begin{center} 
\includegraphics[scale=.35]{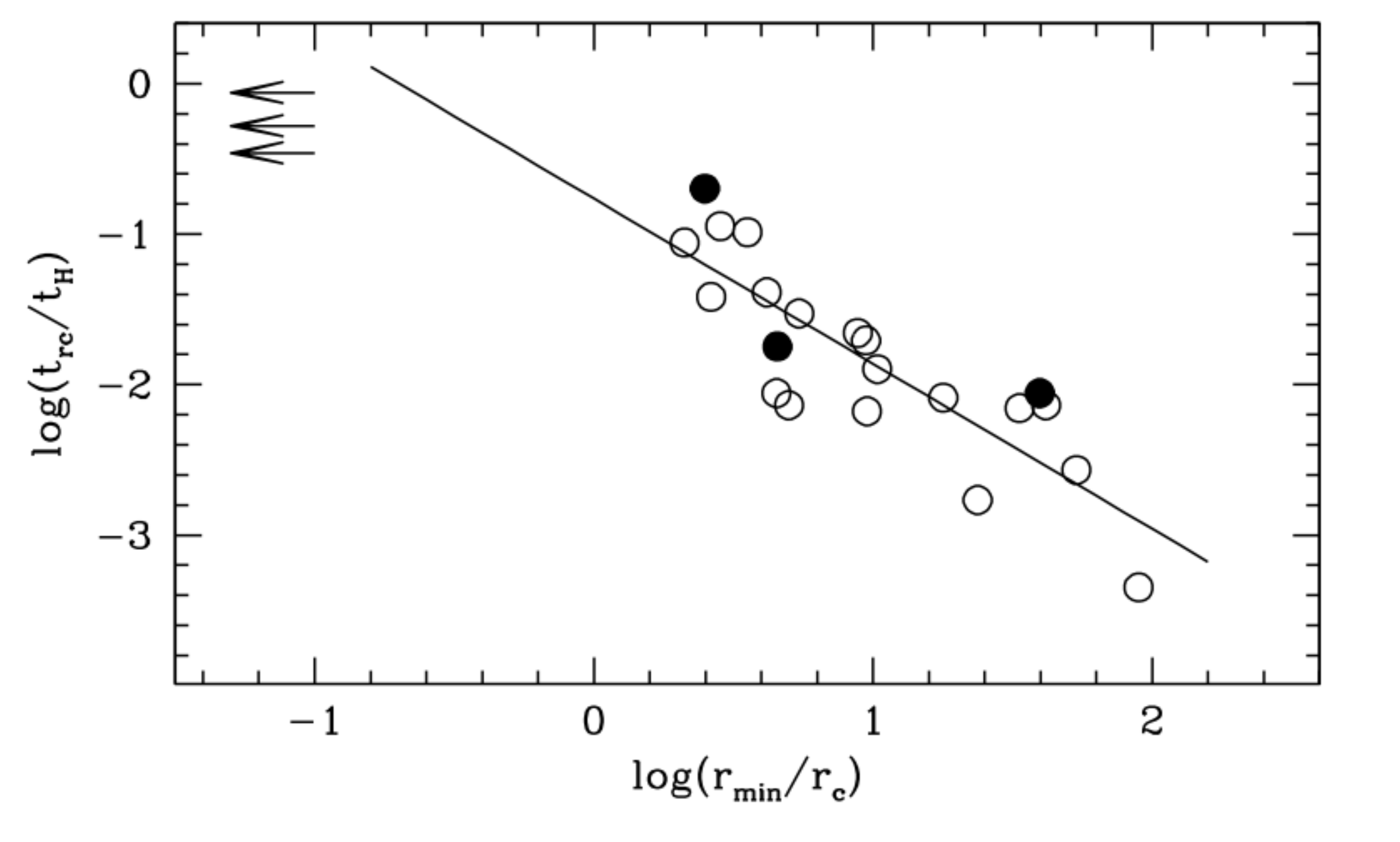}
\caption{\footnotesize Core relaxation time (normalized to the Hubble time $t_H$) as
  a function of the time hand of the proposed \emph{dynamical clock}
  ($r_{\rm min}$, in units of the core radius). Dynamically young
  systems (\emph{Family I}) show no minimum and are plotted as
  lower-limit arrows at $r_{\rm min}/r_c=0.1$.  For dynamically old
  clusters (\emph{Family III}), the distance of the
  farthest radial bin where no BSSs are observed has been adopted as
  $r_{\rm min}$. As expected for a meaningful clock, a tight
  anticorrelation is found: clusters with relaxation times of the
  order of the age of the Universe show no signs of BSS segregation
  (hence their BSS radial distribution is flat and $r_{\rm min}$ is
  not definable; see Fig. \ref{dyncl1}), whereas for decreasing
  relaxation times the radial position of the minimum increases
  progressively.  The solid line correspond to the best-fit relations. 
  Empty circles are data from F12, filled circles are three clusters (NGC362, NGC5824 and NGC5466)
  published after F12.}
\label{dynclock}
\end{center}
\end{figure}

\section{Setting the dynamical clock for stellar systems}
\label{sec:dyn_clock}
In order to perform meaningful comparisons among different clusters
the first step is the definition of an appropriate BSS specific frequency.
\cite[Ferraro et al (1993)]{fer93}
introduced the ``double normalized ratio'', defined as:
\begin{equation}
 R_{\rm BSS} = {{(N_{\rm BSS}/N_{\rm BSS}^{\rm tot})} \over {(L^{\rm
       sampled}/L_{tot}^{\rm sampled})}},
\label{eq:rbss}
\end{equation}
where $N_{\rm BSS}$ is the number of BSSs counted in a given cluster
region, $N_{\rm BSS}^{\rm tot}$ is the total number of BSSs observed,
and $L^{\rm sampled}/L_{tot}^{\rm sampled}$ is the fraction of light
sampled in the same region, with respect to the total measured
luminosity. The same ratio can be defined for any post-MS population.
Theoretical arguments (\cite[Renzini \& Fusi Pecci 1988]{renzfusi88}) demonstrate that the double
normalized ratio is equal to unity for any population (such as red
giant branch and horizontal branch stars, RGB and HB respectively)
whose radial distribution follows that of the cluster
luminosity. 
F12 presented a comparison of the BSS radial distribution of 21 GCs
  with very different structural properties (hence possibly at
different stages of their dynamical evolution), but with nearly the
same chronological age  (12-13 Gry; \cite[Mar{\'{\i}}n-Franch et al.2009]{marinfranch09}), with the
only exception of Palomar 14 which formed $\sim 10.5$ Gyr ago
\cite[Dotter et al.(2008)]{dotter08}.  F12  showed that once the radial distance from
the centre is expressed in units of the core radius (thus to allow a
meaningful comparison among the clusters), GCs can be efficiently
grouped on the basis of the shape of their BSS radial distribution,
and at least three families can be defined:
\begin{itemize}
\item \emph{Family I --} the radial distribution of the BSS double
  normalized ratio ($R_{\rm BSS}$) is fully consistent with that of
  the reference population ($R_{\rm pop}$) over the entire cluster
  extension (see Figure \ref{dyncl1});
\item \emph{Family II --} the distribution of $R_{\rm BSS}$ is
  incompatible with that of $R_{\rm pop}$, showing a significant
  bimodality, with a central peak and an external upturn. At
  intermediate radii a minimum is evident and its position ($r_{\rm
    min}$) can be clearly defined for each sub-group (see Figure
  \ref{dyncl2});
\item \emph{Family III --} the radial distribution of $R_{\rm BSS}$ is
  still incompatible with that of the reference population,
  showing a well defined central peak with no external upturn (see
  Figure \ref{dyncl3}).
\end{itemize}

  Previous
preliminary analysis (\cite[Mapelli et al. (2004)]{mapelli04}, \cite[Mapelli et al. (2006)]{mapelli06},
\cite[Lanzoni et al. 2007a]{lan07}) of a few clusters indicated that BSSs
generated by stellar collisions mainly/only contribute to the central
peak of the distribution, while the portion beyond the observed
minimum is populated by MT-BSSs which are evolving in isolation in the
cluster outskirt and have not yet suffered the effects of dynamical
friction.  Overall, the BSS radial
distribution is primarily modelled by the long-term effect of
dynamical friction acting on the cluster binary population (and its
progeny) since the early stages of cluster evolution. In fact, what we
call MT-BSS today is the by-product of the evolution of a $\sim 1.2
M_\odot$ binary that has been orbiting the cluster and suffering the
effects of dynamical friction for a significant fraction of the
cluster lifetime.  The efficiency of dynamical friction decreases for
increasing radial distance from the centre, as a function of the local
velocity dispersion and mass density.  Hence, dynamical friction first
segregates (heavy) objects orbiting close to the centre and produces a
central peak in their radial distribution. As the time goes on, the
effect extends to larger and larger distances, thus yielding to a
region devoid of these stars (i.e., a dip in their radial
distribution) that progressively propagates outward.  Simple
analytical estimate of the radial position of this dip turned out to
be in excellent agreement with the position of the minimum in the
\emph{observed} BSS radial distributions ($r_{\rm min}$), despite a
number of crude approximation  (see, e.g. \cite[Mapelli et al. 2006]{mapelli06}).
Moreover, a progressive outward drift of $r_{\rm min}$ as a function
of time is confirmed by the results of direct N-body simulations that
follow the evolution of $\sim 1.2 M_\odot$ objects within a reference
cluster over a significant fraction of its 
lifetime (\cite[Miocchi et al. 2015]{miocchi15}).

\begin{figure}[!t]
\begin{center} 
\includegraphics[scale=.55]{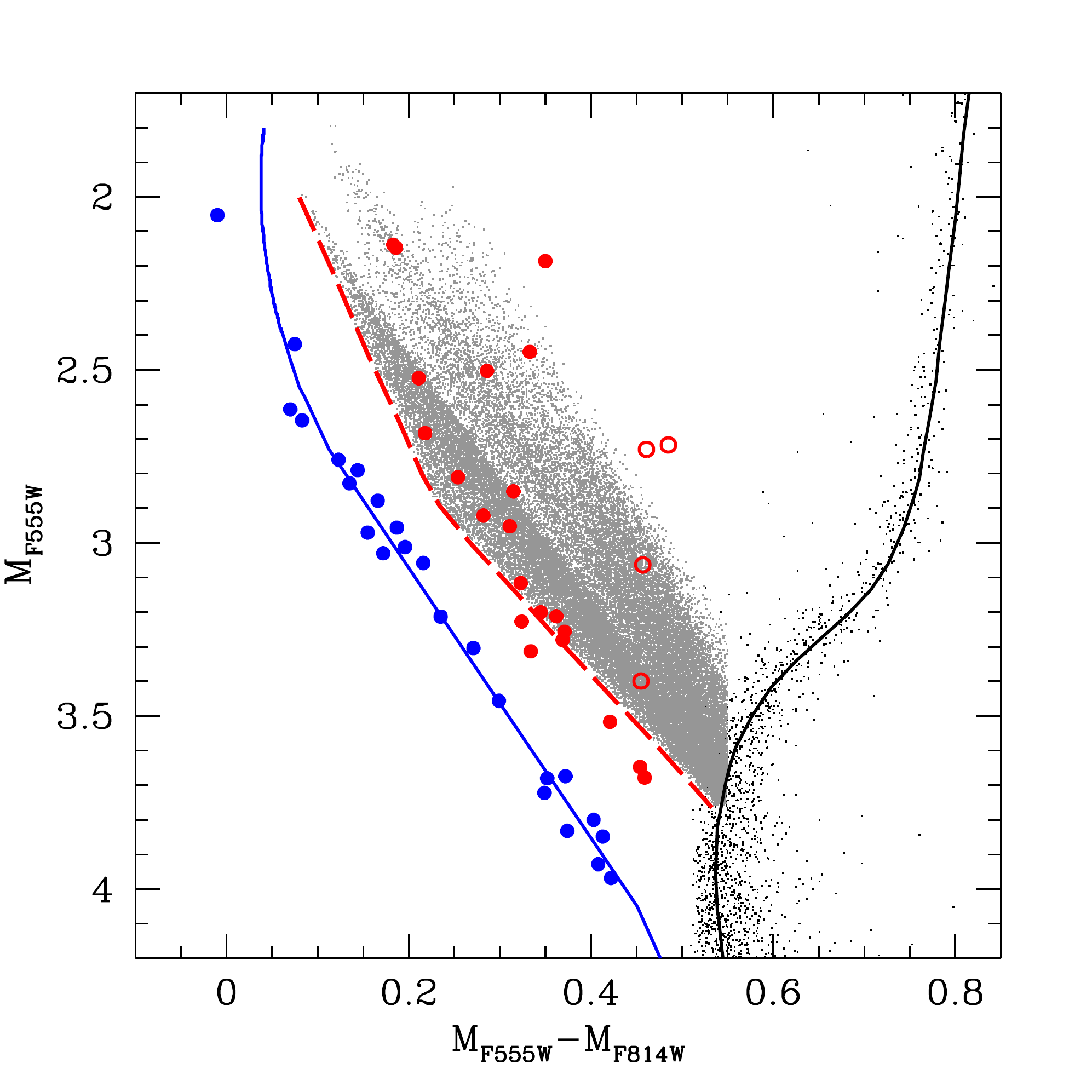}
\caption{\footnotesize Distribution of the synthetic
      MT-binaries (grey dots) bluer than the cluster MS turnoff, in
      the absolute CMD of M30. The dashed line marks the
      ``low-luminosity boundary'' of the distribution. The BSSs
      observed by F09 along the red and the blue sequences are shown
      as solid circles. The positions of 4
      additional red-BSSs, not considered in F09, are also shown as
      empty circles.  Clearly, the distribution of synthetic
      BSSs well samples the location of the observed red-BSSs, and its
      low-luminosity boundary nicely follows the red-BSS sequence. The
       solid line is the collisional 2 Gyr-isochrone by \cite[Sills et al (2009)]{sills09}, which has 
       been found to nicley
      fit the observed blue-BSS sequence.}
\label{m30}
\end{center}
\end{figure}
 
In light of these considerations, the three families defined in
Figs. \ref{dyncl1}--\ref{dyncl3} correspond to GCs of increasing
dynamical ages. Hence, the shape of the BSS radial distribution turns
out to be a powerful dynamical-age indicator. A flat BSS radial
distribution (consistent with that of the reference population; see
{\it Family I} in Fig. \ref{dyncl1}) indicates that dynamical friction
has not played a major role yet even in the innermost regions, and the
cluster is still dynamically young. This interpretation is confirmed
by the absence of statistically significant dips in the BSS
distributions observed in dwarf spheroidal galaxies 
(\cite[Mapelli et al. 2009]{mapelli09}; \cite[Monelli et al 2012]{monelli12}): these
are, in fact, collisionless systems where dynamical friction is
expected to be highly inefficient.  In more evolved clusters ({\it
  Family II}), dynamical friction starts to be effective and to
segregate BSSs that are orbiting at distances still relatively close
to the centre: as a consequence, a peak in the centre and a minimum at
small radii appear in the distribution, while the most remote BSSs are
not yet affected by the action of dynamical friction (this generates
the rising branch of the observed bimodal BSS distributions; see upper
panel in Fig. \ref{dyncl2}).  Since the action of dynamical friction
progressively extends to larger and larger distances from the centre,
the dip of the distribution progressively moves outward (as seen in
the different groups of {\it Family II} clusters; Fig. \ref{dyncl2},
panels from top to bottom).  In highly evolved systems dynamical
friction already affected even the most remote BSSs, which started to
gradually drift toward the centre: as a consequence, the external
rising branch of the radial distribution disappears (as observed for
{\it Family III} clusters in Fig. \ref{dyncl3}). All GCs with a
single-peak BSS distribution can therefore be classified as
``dynamically old''.

Interestingly, this latter class includes M30, a system that already suffered core collapse which
is considered as a typical symptom of extreme dynamical evolution
(\cite[Meylan \& Heggie {1997}]{mh97}).  The proposed classification is also
able to shed light on a number of controversial cases debated into the
literature.  In fact, M4 turns out to have an intermediate dynamical
age, at odds with previous studies suggesting that it might be in a
PCC state (\cite[Heggie \& Giersz 2008]{heggie08m4}). On the other hand, NGC 6752 turns out
to be in a quite advanced state of dynamical evolution, in agreement
with its observed double King profile indicating that the cluster core
is detaching from the rest of the cluster structure (\cite[Ferraro et al 2003b]{fe036752}).
Finally this approach might provide the key to discriminate between a
central density cusp due to core collapse (as for M30) and that due to
the presence of an exceptional concentration of dark massive objects
(neutron stars and/or the still elusive intermediate-mass black
  holes; (see the case of NGC 6388, \cite[Lanzoni et al. 2007c]{lan07imbh}).

The quantization in distinct age-families is of course an
over-simplification, while the position of $r_{\rm min}$ is found to
vary with continuity as a sort of clock time-hand.  This allowed F12
to define the first empirical clock able to measure the dynamical age
of a stellar system from pure observational quantities (the {\it
  "dynamical clock"}): as the engine of a chronometer advances the
clock hand to measure the time flow, in a similar way the progressive
sedimentation of BSSs towards the cluster centre moves $r_{\rm min}$
outward, thus marking its dynamical age.  This is indeed confirmed by
the tight correlations found between the clock-hand ($r_{\rm min}$)
and the central and half-mass relaxation times ($t_{\rm rc}$ and
$t_{\rm rh}$, respectively), which are commonly used to measure the
cluster dynamical evolution time-scales.  
The correlation found with $t_{\rm rc}$ is shown in Figure \ref{dynclock}.
Note that, while $t_{\rm rc}$
and $t_{\rm rh}$ provide an indication of the relaxation timescales at
specific radial distances from the cluster centre ($r_{\rm c}$ and
$r_{\rm h}$, respectively), the dynamical clock here defined provides instead
a measure of the global dynamical evolution of the systems, because
the BSS radial distribution simultaneously probes all distances from
the cluster centre.

\section{The discovery of the double BSS sequence}
\label{sec:double}

While the proposed BSS formation mechanisms could be separately at
work in clusters with different densities (\cite[Ferraro et al 1995]{fer95}, \cite[Ferraro et al. 2006b]{fer06co},
\cite[Lovisi et al. 2013]{lovi13}) a few
pieces of evidence are now emerging suggesting that they could also
act simultaneously within the same cluster. Indeed the discovery of a
double BSS sequence in M30 (\cite[Ferraro et al 2009b]{fer09m30}, hereafter F09) indicates
that this can be the case: two well distinct BSS sequences, almost parallel and
similarly populated, have been found in the CMD of M30 (F09).  Similar
  features have been detected also in NGC~362 by \cite[Dalessandro et
  al. (2013)]{dale13} and NGC~1261 by \cite[Simunovic et al. (2014)]{simu14}.
  
Such a   
 discovery has potentially opened the possibility to
photometrically distinguish collisional BSSs from MT BSSs in the same
cluster. In fact, the blue-BSS sequence
observed in M30 is nicely reproduced by collisional isochrones (\cite[Sills
et al. 2009]{sills09}) with ages of 1-2 Gyr.  In contrast, the red-BSS
population is far too red to be consistent with collisional isochrones
of any age. F09 suggested that the location of the red-BSS population
in the CMD correspond to the extrapolation of the   the ``low-luminosity
boundary'' outlined by the MT binary populations simulated by \cite[Tian et
al. (2006)]{tian06} for the open cluster M67. Indeed, by  
  using binary evolution models specifically computed for M30, \cite[Xin et al. (2015)]{xin15} 
demonstrated that the  distribution of synthetic MT-BSSs   
 in the CMD  nicely matches the observed red BSS
sequence (see Figure \ref{m30}), thus providing strong support to the MT origin
for these stars. In addition the CMD distribution of synthetic MT-BSSs never
attains the observed location of the blue BSS sequence, thus
reinforcing the hypothesis that the latter formed through a different
channel (likely collisions);

\vskip5truemm
Most of the results discussed in this talk have been obtained within 
the project Cosmic-Lab (PI: Ferraro, see http:://www.cosmic-lab.eu), a 5-year 
project funded by the European European Research Council under the 2010 Advanced 
Grant call (contract ERC-2010-AdG-267675). I warmly thank the other team members 
involved in this research: Barbara Lanzoni, Emanuele Dalessandro, Alessio Mucciarelli, 
Yu Xin, Licai Deng, Giacomo Beccari, Paolo Miocchi, Mario Pasquato.


\vskip19truemm

\begin{thebibliography}{}
 
\bibitem[Alessandrini et al.(2014)]{ale14} Alessandrini, E., Lanzoni, B., Miocchi, P., Ciotti, L., Ferraro, F.~R,. 2014, \textit{ApJ}, 795, 169 
\bibitem[Bailyn (1992)]{bai92} Bailyn, C.  1992, \textit{ApJ}, 392, 519
\bibitem[Bailyn (1995)]{bailyn95} Bailyn, C.~D.,1995, \textit{ARAA}, 33, 133
\bibitem[Beccari et al.(2011)]{bec11} Beccari, G., Sollima,  A., Ferraro, F.~R., et al.,  2011, \textit{ApJ}, 737, L3
\bibitem[Beccari et al. (2013)]{bec13} Beccari, G., Dalessandro, E., Lanzoni, B., et al., 2013, \textit{ApJ}, 776, 60
\bibitem[Bellazzini et al.(1995)]{bellazzini95} Bellazzini, M., Pasquali, A., Federici, L., Ferraro, F.~R., \& Pecci, F.~F. 1995, \textit{ApJ}, 439, 687
\bibitem[Benz \& Hills (1987)]{benz87} Benz, W., \& Hills, J. G., 1987, \textit{ApJ}, 323, 614
\bibitem[Buonanno et al.(1994)]{buo94} Buonanno, R., Corsi, C.~E., Buzzoni, A., et al.,1994, \textit{A\&A}, 290, 69 
\bibitem[Chen \& Han (2009)]{chen09} Chen, X. F., \& Han, Z. W.,2009, \textit{MNRAS}, 395, 1822
\bibitem[Contreras Ramos et al.(2012)]{contre12} Contreras Ramos, R., Ferraro, F.~R., Dalessandro, E., Lanzoni, B., \& Rood, R.~T.,2012, \textit{ApJ}, 748, 91
\bibitem[Dalessandro et al. (2008a)]{dale2419} Dalessandro, E., Lanzoni, B., Ferraro, F. R., Vespe, F., Bellazzini, M., \& Rood, R. T.,2008a, \textit{ApJ}, 681, 311
\bibitem[Dalessandro et al. (2008b)]{dale6388} Dalessandro, E., Lanzoni, B., Ferraro, F. R., Rood, R. T., Milone, A., Piotto, G., \& Valenti, E.,2008b, \textit{ApJ}, 677, 1069
\bibitem[Dalessandro et al. (2013)]{dale13} Dalessandro, E., Ferraro, F. R., Massari, D., Lanzoni, B., Miocchi, P., Beccari, G., Bellini, A., Sills, A., Sigurdsson, S., Mucciarelli, A., \& Lovisi, L.,2013, \textit{ApJ}, 778, 135
\bibitem[De~Marco et al. (2004)]{de04} De~Marco, O., Lanz, T., Ouellette, J. A., Zurek, D., \& Shara, M. M.,2004, \textit{ApJ}, 606, L151
\bibitem[Davies et al.(2004)]{davies04} Davies, M.~B., Piotto, G., \& de Angeli, F.,2004, \textit{MNRAS}, 349, 129
\bibitem[Dotter et al.(2008)]{dotter08} Dotter, A., Sarajedini, A., \&  Yang, S.-C.,2008, \textit{AJ}, 136, 1407
\bibitem[Ferraro et al. (1993)]{fer93} Ferraro F. R., Pecci F. F.,Cacciari C., Corsi C., Buonanno R., Fahlman G. G., Richer H. B.,1993,\textit{AJ}, 106, 2324
\bibitem[Ferraro et al. (1995)]{fer95} Ferraro, F. R., Fusi Pecci, F.,  \& Bellazzini, M.,1995, \textit{A\&A}, 294, 80
\bibitem[Ferraro et al. (1997)]{fer97} Ferraro, F. R., et al.,Paltrinieri, B., Fusi Pecci, F., Cacciari, C., Dorman, B., Rood, R. T., Buonanno, R., Corsi, C. E., Burgarella, D., \& Laget, M., 1997,\textit{A\&A}, 324, 915 
\bibitem[Ferraro et al. (1999a)]{fer99m80} Ferraro, F. R., Paltrinieri, B., Rood, R. T., \& Dorman, B.,1999a, \textit{ApJ}, 522, 983
\bibitem[Ferraro et al. (1999b)]{fer99cat} Ferraro, F.~R., Messineo, M., Fusi Pecci, F., et al.,1999b, \textit{AJ}, 118, 1738 
\bibitem[Ferraro et al.(2001)]{fer01} Ferraro, F.~R., D'Amico, N.,  Possenti, A., Mignani, R.~P., \& Paltrinieri, B.,2001, \textit{ApJ}, 561,337
\bibitem[Ferraro et al. (2003a)]{fer03six} Ferraro, F. R., Sills, A., Rood, R. T., Paltrinieri, B., \& Buonanno, R.,2003a, \textit{ApJ}, 588, 464
\bibitem[Ferraro et al. (2003b)]{fer036752} Ferraro, F.~R., Possenti, A., Sabbi, E., et al.,2003b, \textit{ApJ}, 595, 179 
\bibitem[Ferraro et al. (2004)]{fer04} Ferraro, F. R., Beccari, G., Rood, R. T., Bellazzini, M., Sills, A., \& Sabbi, E. ,2004, \textit{ApJ}, 603, 127
\bibitem[Ferraro et al. (2006a)]{fer06ome} Ferraro, F. R., Sollima, A., Rood, R. T., Origlia, L., Pancino, E., \& Bellazzini, M.,2006a, \textit{ApJ}, 638, 433
\bibitem[Ferraro et al. (2006b)]{fer06co} Ferraro, F. R., Sabbi, E., Gratton, R., et al., 2006b, \textit{ApJ}, 647, L53
\bibitem[Ferraro et al.(2009a)]{fer09ter} Ferraro, F.~R., Dalessandro, E., Mucciarelli, A., et al.,2009a, \textit{Nature}, 462, 483 
\bibitem[Ferraro et al. (2009b)]{fer09m30} Ferraro, F. R., Beccari, G., Dalessandro, E., et al.,2009b, \textit{Nature}, 462, 1028 (F09)
\bibitem[Ferraro et al. (2012)]{fer12} Ferraro, F.~R.,  Lanzoni, B., Dalessandro, E., et al.,2012, \textit{Nature}, 492, 393, (F12) 
\bibitem[Fiorentino et al.(2014)]{fiore14} Fiorentino, G., Lanzoni, B., Dalessandro, E., et al.,2014, \textit{ApJ}, 783, 34 
\bibitem[Fusi Pecci et al. (1992)]{fp92} Fusi Pecci, F., Ferraro, F. R., Corsi, C. E., Cacciari, C., \& Buonanno, R.,1992, \textit{ApJ}, 104, 1831
\bibitem[Gilliland et al.(1998)]{gilli98} Gilliland, R.~L., Bono,G., Edmonds, P.~D., et al.,1998, \textit{ApJ}, 507, 818
\bibitem[Glebbeek et al. (2008)]{gle08} Glebbeek, E., Pols, O. R., \& Hurley, J. R.,2008, \textit{A\&A}, 488, 1007
\bibitem[Gosnell et al. (2014)]{Gos14} Gosnell, N. M., Mathieu, R. D., Geller, A. M., Sills, A., Leigh, N., \& Knigge, C., 2014, \textit{ApJ}, 783, L8
\bibitem[Heggie \& Giersz(2008)]{heggie08m4} Heggie, D.~C., \& Giersz, M.,2008, \textit{MNRAS}, 389, 1858
\bibitem[Hut et al. (1992)]{hut92} Hut, P., McMillan, S., \& Romani, R.~W.,1992, \textit{ApJ}, 389, 527
\bibitem[Hills \& Day (1976)]{hills76} Hills, J., \& Day, C.,1976, Astron. Lett., 17, 87
\bibitem[Hurley et al. (2001)]{hur01} Hurley, J. R., Tout, C. A., Aarseth, S. J., \& Pols, O. R.,2001, \textit{MNRAS}, 323, 630
\bibitem[Knigge et al. (2009)]{knigge2009} Knigge, C., Leigh, N., \& Sills, A.,2009, \textit{Nature}, 457, 288
\bibitem[Lanzoni et al. (2007a)]{lan07} Lanzoni, B., Dalessandro, E., Perina, S., Ferraro, F. R., Rood, R. T., \&  Sollima, A. ,2007a, \textit{ApJ}, 670, 1065
\bibitem[Lanzoni et al. (2007b)]{lan071904} Lanzoni, B., Sanna, N., Ferraro, F.~R., et al.,2007b, \textit{ApJ}, 663, 1040
\bibitem[Lanzoni et al.(2007c)]{lan07imbh} Lanzoni, B., Dalessandro, E., Ferraro, F.~R., et al.,2007c, \textit{ApJ}, 668, L139 
\bibitem[Lanzoni et al.(2010)]{lan10} Lanzoni, B., Ferraro, F.~R., Dalessandro, E., et al.,2010, \textit{ApJ}, 717, 653 
\bibitem[Leigh et al.(2007)]{leigh07}  Leigh, N., Sills, A., \& Knigge, C.,2007, \textit{ApJ}, 661, 210 
\bibitem[Leigh et al. (2013)]{leigh13} Leigh, N., Knigge, C., Sills, A., Perets, H.B., Sarajedini, A., \& Glebbeek, E.,2013, \textit{MNRAS}, 428, 897
\bibitem[Leonard (1989)]{leo89} Lenoard, P.~J.~T.,1989, \textit{AJ}, 98, 217
\bibitem[Leonard  \& Linnell(1992)]{leli92} Leonard, P.~J.~T., \& Linnell, A.~P.,1992, \textit{AJ}, 103, 1928 
\bibitem[Lombardi et al. (1995)]{lom95} Lombardi, Jr., J. C., Rasio, F. A., \& Shapiro, S. L.,1995, \textit{ApJ}, 445, L117
\bibitem[Lombardi et al. (2002)]{lom02} Lombardi, Jr., J. C., Warren, J. S., Rasio, F. A., Sills, A., \& Warren, A. R., 2002, \textit{ApJ}, 568, 939
\bibitem[Lovisi et al. (2013)]{lovi13} Lovisi, L., Mucciarelli, A., Lanzoni, B., Ferraro, F. R., Dalessandro, E., \& Monaco, L., 2013, \textit{ApJ}, 772, 148
\bibitem[Mapelli et al. (2004)]{mapelli04} Mapelli, M., Sigurdsson, S., Colpi, M., Ferraro, F. R., Possenti, A., Rood, R. T., Sills,A., \& Beccari, G.,2004, \textit{ApJ}, 605, L29
\bibitem[Mapelli et al. (2006)]{mapelli06} Mapelli, M., Sigurdsson,S., Ferraro, F. R., Colpi, M., Possenti, A., \& Lanzoni, B.,2006,\textit{MNRAS}, 373, 361
\bibitem[Mapelli et al.(2009)]{mapelli09} Mapelli, M., Ripamonti, E., Battaglia, G., et al.,2009, \textit{MNRAS}, 396, 1771
\bibitem[Miocchi et al.(2015)]{miocchi15} Miocchi, P., Pasquato, M., Lanzoni, B., et al.\ 2015, \textit{ApJ}, 799, 44 
\bibitem[Mar{\'{\i}}n-Franch et al.(2009)]{marinfranch09} Mar{\'{\i}}n-Franch, A., Aparicio, A., Piotto, G., et al.,2009,\textit{ApJ}, 694, 1498
\bibitem[Mathieu \& Geller(2009)]{mathieu09} Mathieu, R.~D., \& Geller, A.~M.,2009, \textit{Nature}, 462, 1032
\bibitem[McCrea (1964)]{mc64} McCrea, W. H.,1964, \textit{MNRAS}, 128, 147
\bibitem[Meylan \& Heggie {1997}]{mh97} Meylan, G., \& Heggie, D.C. 1997, \textit{ARAA}, 8, 1
\bibitem[Monelli et al.(2012)]{monelli12} Monelli, M., Cassisi, S., Mapelli, M., et al.,2012, \textit{ApJ}, 744, 157
\bibitem[Ouellette \& Pritchet (1998)]{oue98} Ouellette, J. A., \& Pritchet, C. J., 1998, \textit{AJ}, 115, 2539
\bibitem[Paresce et al. (1992)]{paresce92} Paresce, F., de Marchi, G.,\& Ferraro, F.~R.,1992, \textit{Nature}, 360, 46
\bibitem[Pooley \& Hut(2006)]{pooleyhut06} Pooley, D., \& Hut, P.,2006, \textit{ApJ}, 646, L143
\bibitem[Piotto et al. (2004)]{pio04} Piotto, G., et al.,2004, \textit{ApJ}, 604, L109
\bibitem[Preston \& Sneden(2000)]{pre00} Preston, G.~W., \& Sneden, C.,2000, \textit{AJ}, 120, 1014 
\bibitem[Ransom et al.(2005)]{ransom05} Ransom, S.~M., Hessels, J.~W.~T., Stairs, I.~H., et al.,2005, Science, 307, 892
\bibitem[Renzini \& Fusi Pecci(1988)]{renzfusi88} Renzini, A., \& Fusi Pecci, F.,1988, \textit{ARAA}, 26, 199
\bibitem[Shara et al. (1997)]{shara97} Shara, M. M., Saffer, R. A., \& Livio, M.,1997, \textit{ApJ}, 489, L59
\bibitem[Sills et al.(2009)]{sills09} Sills, A., Karakas, A., \& Lattanzio, J.,2009, \textit{ApJ}, 692, 1411
\bibitem[Simunovic et al.(2014)]{simu14} Simunovic, M., Puzia, T.~H., \& Sills, A.,2014, \textit{ApJ}, 795, L10 
\bibitem[Sollima et al. (2008)]{soll08} Sollima, A., Lanzoni, B., Beccari, G., Ferraro, F. R., \& Fusi~Pecci, F., 2008, \textit{A\&A}, 481, 701
\bibitem[Tian et al. (2006)]{tian06} Tian, B., Deng, L.-C., Han, Z.-W., \& Zhang, X.-B., 2006, \textit{A\&A}, 455, 247
\bibitem[Xin et al.(2015)]{xin15} Xin, Y., Ferraro, F.~R., Lu, P., et al.\ 2015, arXiv:1501.01358 
\end{thebibliography}
\end{document}